\documentclass[prl,twocolumn,nofootinbib,preprintnumbers,amssymb,amsfonts,amsmath,superscriptaddress,showpacs,hyperref]{revtex4-1}

\usepackage{amsmath,array,amssymb}
\usepackage{graphicx}
 \usepackage{braket}
 
\usepackage{color}

\begin{document}

\title{Optical manipulation of a magnon-photon hybrid system}%

\author{C. Braggio}
\email[Electronic address: ]{caterina.braggio@unipd.it}
\affiliation{Dip. di Fisica e Astronomia and INFN, Sez di Padova, Via F. Marzolo 8, I-35131 Padova, Italy}

\author{G. Carugno}
\affiliation{Dip. di Fisica e Astronomia and INFN, Sez di Padova, Via F. Marzolo 8, I-35131 Padova, Italy}

\author{M. Guarise}
\affiliation{Dip. di Fisica e Astronomia and INFN, Sez di Padova, Via F. Marzolo 8, I-35131 Padova, Italy}

\author{A. Ortolan}
\email[Electronic address: ]{antonello.ortolan@lnl.infn.it}
\affiliation{INFN, Laboratori Nazionali di Legnaro, Viale dell'Universit\`a 2, I-35020 Legnaro, Italy}

\author{G. Ruoso}
\affiliation{INFN, Laboratori Nazionali di Legnaro, Viale dell'Universit\`a 2, I-35020 Legnaro, Italy}


\begin{abstract}

We demonstrate an all-optical method 
for manipulating the magnetization in a 1--mm YIG (yttrium-iron-garnet) sphere placed in a $\sim0.17$\,T uniform magnetic field. An harmonic of the frequency comb delivered by a multi-GHz infrared laser source is tuned to the Larmor frequency of the YIG sphere to drive magnetization oscillations, which in turn give rise to a radiation field 
used to thoroughly investigate the phenomenon. 
 The radiation damping issue that occurs at high frequency and in the presence of highly magnetizated materials, has been overcome by exploiting magnon-photon strong coupling regime in microwave cavities. 
 Our findings demonstrate an effective technique for ultrafast control of the magnetization vector in optomagnetic materials via polarization rotation and intensity modulation of an incident laser beam. We eventually get a second-order susceptibility value of $\sim10^{-7}$ cm$^2$/MW for single crystal YIG.
 
\end{abstract}

\maketitle
{\em Introduction.}---Nonthermal control of spins by short laser pulses is one of the preferable means to achieve ultrafast control of the magnetization in magnetic materials \cite[see][and references therein]{Kirilyuk:2010}, representing a breakthrough in potential applications ranging from high density magnetic data storage \cite{Stanciu:2007vn}, spintronics \cite{Li:2013zr}, to quantum information processing \cite{Xiang:2013kx,Zhang:2015uq}. 
One of the most interesting opto-magnetic mechanisms that allows for coherent control of the magnetization in materials is the Inverse Faraday (IF) effect, a Raman-like coherent scattering process that entails the generation of a magnetic excitation (i.e. magnon) in a medium undergoing the action of high--intensity optical pulses. As it does not require absorption and takes place on a femtosecond time scale, it stands out as a promising mechanism to control the magnetization at high speeds. This principle has been successfully applied in dysprosium orthoferrite (DyFeO$_3$) with isolated femtosecond laser pulses, that act as magnetic field pulses of a 0.3\,T amplitude \cite{Kimel:2005}. By way of the IF effect, vector control of magnetization in another antiferromagnetic crystal was also demonstrated by varying the delay between pairs of polarization-twisted ultrashort optical pulses \cite{Kanda:2011}. 
In this work we introduce a new approach in opto--magnetism based on multi-gigahertz repetition rate lasers with optical carrier $f_0$ \cite{Keller:2003}.  The power spectrum of such mode-locked laser sources, as detected by ultrafast photodiodes, is a frequency comb that consists of several har\-mo\-nics $nf_r$, where $f_r$ is the repetition rate and $n$ is a small number.

Their gaussian envelope is determined by the optical pulse temporal profile \cite{Cundiff:2003}. For example, our 4.6\,GHz passively mode-locked oscillator delivers $\sim$10\,ps-duration pulses that give rise to a frequency comb up to 100\,GHz, and the first three harmonics have approximately the same amplitude. 
In principle, any harmonic of the comb can coherently drive the magnetization in the steady state through the process described in the present work, provided it is tuned to electron spin resonances (ESR) of the magnetized material.

 We study the spin dy\-na\-mics in a hybridized system which consists of two strongly coupled oscillators, i.e. a microwave cavity mode and a magnetostatic mode related to ferromagnetic resonance (FMR) with uniform precession \cite{Kittel:2005} of a single crystal yttrium-iron garnet Y$_3$Fe$_5$O$_{12}$ sphere. Magnetic garnets \cite{Hansteen:2005fk, Satoh:2012fk} represent the ideal materials for such investigations for several reasons, including the possibility to realize large magneto-optical effects due to their strong spin-orbit coupling and intrinsically low magnetic damping \cite{Serga:06, Satoh:2012fk, Kajiwara:2010fk}. 

We succeed to optically drive the precession of the spins electro-dynamically coupled to the cavity photons with the first harmonic of the train of pulses at $f_r$ tuned to one of the hybrid system's resonant frequencies. The process gives rise to a microwave field that is measured with a loop antenna critically coupled to the cavity mode. In this way, we have 
 identified a new observable for  the spin precession to explore opto-magnetic phenomena. So far, experiments in this field were performed with pump-probe apparatus based on femtosecond lasers \cite{Bossini:2016ys,Kanda:2011,Satoh:2012fk,Kirilyuk:2010}.  \\
\indent As it is well known, at very high values of frequency and magnetization, the energy radiated from oscillating magnetization through magnetic dipole radiation  can be an issue for the dynamic control of the magnetization.  For instance, in a polarized 1--mm YIG sphere with linear susceptibility $\chi\sim30$, the onset of radiation damping occurs at $\sim 10\,$GHz \cite{Bloembergen:1954fk}. 
However, radiation damping mechanism can be conveniently suppressed  in the microwave cavity-QED and strong coupling regimes \cite{Scully:1997, Bloembergen:1954fk}, as we detail in the following for hybridized systems. Moreover, under  remarkable conditions of hybridization, we get  the control of relevant experimental parameters such as the number of spins, rf absorbed power, and the involved relaxation times. 
System hybridization is however not essential to observe the phenomenon described in the present work. In fact, we succeed in controlling the magnetization also in free space, but under experimental conditions that do not allow for accurate modeling.

\begin{figure}[h!]
\begin{center}
\includegraphics[width=3.4 in]{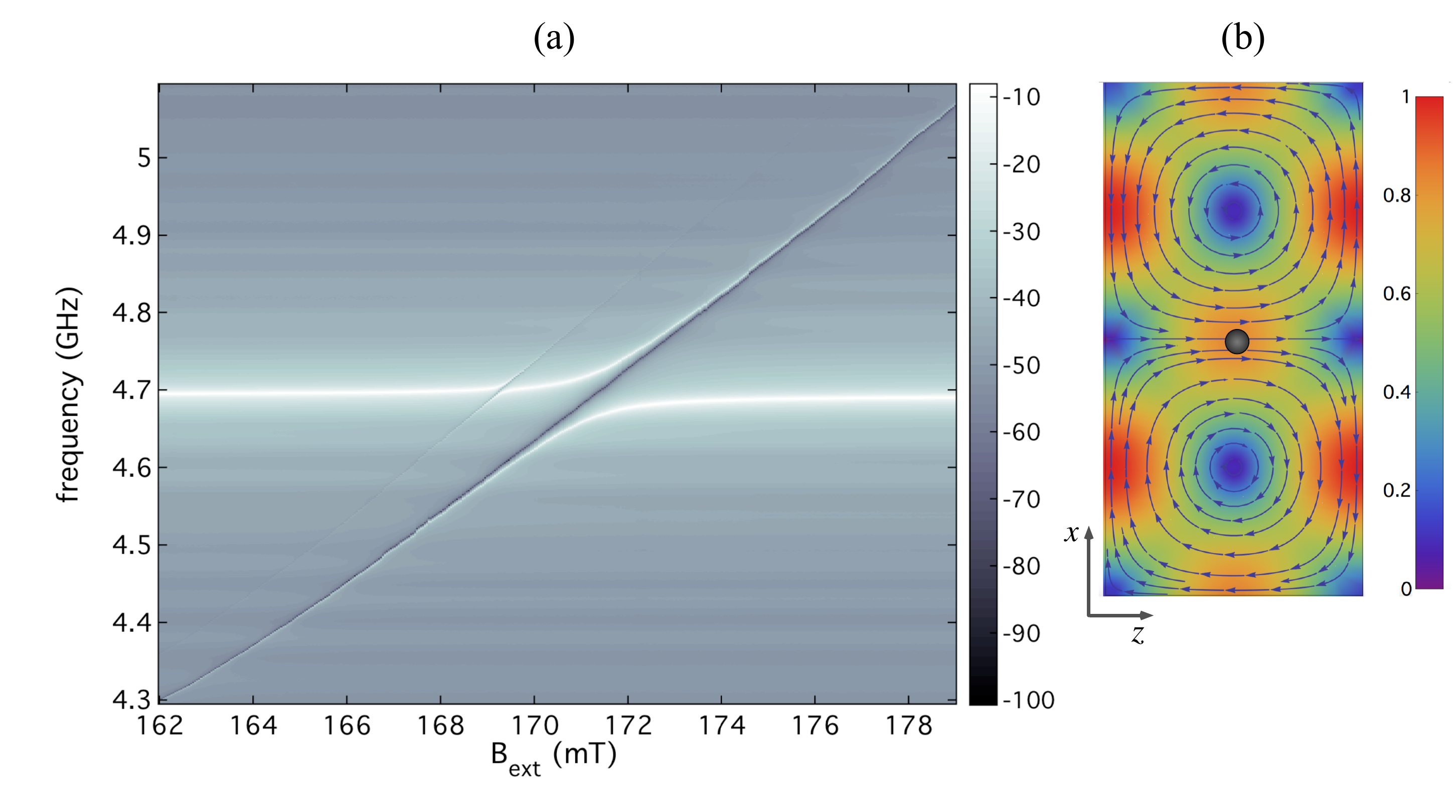}
\caption{Hybridizing magnons and microwave photons. (a) Cavity transmission spectrum measured as a function of the static magnetic field at room temperature. 
 (b)  Simulated magnetic--field distribution of the TE$_{102}$ cavity mode. A static magnetic field B$_{\rm ext}$ is applied normal to the $xz$ plane, and the microwave magnetic field at the YIG sphere (in black and not to scale in the representation) position is perpendicular to the static magnetic field.  The color map represents the amplitude of the cavity magnetic field normalized to its maximum value.
 }
\label{coup}
\end{center}
\end{figure}

{\em Hybridized system characterization.}---
Strong interaction between light (i.e. photons stored in a cavity) and magnetized materials has been accomplished in several experiments that are paving the way toward the development of quantum information technologies \cite{Huebl:2013vn, Zhang:2014, Tabuchi:2014uq}. Hybridization is commonly investigated by measuring the microwave--cavity transmission spectrum as a function of the static magnetic field, as summarized in Fig.\,\ref{coup} for our experimental setup, even though, very recently, some authors have reported electric detection via spin pumping \cite{Bai:2015kx}.
A YIG sphere made by Ferrisphere Inc. with a radius of 1\,mm is mounted at the center of a 3D rectangular microwave cavity with dimensions $98\times42.5\times12.6$\,mm$^3$. The cavity made of oxygen free copper has the TE$_{102}$ mode frequency $\omega_c/2\pi \simeq 4.67$\,GHz, and its internal cavity loss $\kappa_{int}$. This cavity 
has two ports characterized by the coupling coefficients $\kappa_1$ and $\kappa_2$  to the considered cavity mode.  
The sphere is glued to an alumina (aluminum--oxide) rod that identifies the crystal axis $[110]$, perpendicular to the static magnetic field $\mathbf B_{\rm ext}$ ($y$ axis) and parallel to the TE$_{102}$ microwave magnetic field lines lying on the $xz$ plane.  
 Due to the strong coupling between the cavity mode and FMR mode an avoided crossing occurs when their resonant frequencies match.
As derived in the input--output theory context \cite{Scully:1997, Tabuchi:2014uq}, when the static magnetic field is tuned to drive the magnons in resonance with the cavity mode TE$_{102}$, the measured transmission coefficient can be written as 
\begin{equation}
\label{S21}
S_{21}(\omega)=\frac{\sqrt{\kappa_1\kappa_2}}{i(\omega-\omega_c)-\frac{\kappa_c}{2}+\frac{|g_m|^2}{i(\omega-\omega_{\rm FMR})-\gamma_m/2}}\,, 
\end{equation} 
where $\omega_{\rm FMR}$ and $\gamma_m$ are the frequency and linewidth of the FMR mode, $\kappa_c/2\pi=(\kappa_1+\kappa_2+\kappa_{int})/2\pi$ is the total cavity linewidth, and $g_m$ is the coupling strength of the FMR mode to the cavity mode. The latter parameter is proportional to the square root of the number of precessing spins $N_s$, i.e. $g_m=g_0\sqrt{N_s}$, where $g_0=\gamma_{e}\sqrt{\mu_{0}\hbar \omega_{c}/V_{c}}$ is the coupling strength of a single spin to the cavity mode, with $\gamma_{e}=2\pi \times 28\,$GHz/T electron gyromagnetic ratio, $\mu_{0}$ permeability of vacuum and $V_{c}$ is cavity volume. 

As discussed in the seminal paper of Bloembergen and Pound  \cite{Bloembergen:1954fk}, the poles at the anti--crossing point $\omega_{\rm FMR}=\omega_c\equiv\omega_0$ are given by
\begin{equation}
\label{poles}
p_\pm\!=\!i\!\left(\!\omega_0 \! \pm \! \sqrt{|g_m|^2\!-\![(\kappa_c - \gamma_m)/4]^2} \right)\!-\!
\frac{1}{2} \left(\!\frac{k_c}{2}\!+\!\frac{\gamma_m}{2}\! \right)\!,
\end{equation}
and their imaginary and real parts represent the frequencies  
\begin{displaymath}
\omega_{\pm}= \omega_0   \pm  \sqrt{|g_m|^2-[(\kappa_c-\gamma_m)/4]^2}
\end{displaymath}
and the linewidths  $\gamma_{\pm}=1/2(k_c+\gamma_m)$
of the hybridized modes, respectively.  From eq.\,\ref{poles} hybridization clearly occurs only if $|g_m|^2-[(\kappa_c+\gamma_m)/4]^2>0$ and, as a consequence, hybridized modes have the same decay time, independent of the sample or cavity volume and coupling strengths, i.e.
$\bar{\tau}\equiv\tau_{\pm}=(2/\tau_c + 2/\tau_2)^{-1}$,
where $\tau_c=2/\kappa_c$ and $\tau_2$ are the loaded cavity decay time and  the spin--spin relaxation time, respectively. In the absence of hybridization, the term under square root in eq.\,\ref{poles} is negative and thus the poles have the same  
frequency $\omega_0$, with two relaxation times $\tau_c$ and $\tau^{\ast}= (1/\tau_{r} +1/\tau_2)^{-1}$, that correspond to the damping of the cavity mode and magnetization mode in the presence of radiation damping $\tau_r=\kappa_c/2/ |g_m|^2$.

In our experimental apparatus for B$_{\rm ext}\simeq 0.17$\,T we achieve a strong coupling regime with $g_{m}/2\pi=57$\,MHz, 
 thus the involved precessing spins are $N_s\sim 10^{20}$. Along with the mode frequencies $f_+=4.7247$\,GHz and $f_-=4.6677$\,GHz, the fit 
of the measured $S_{21}$ coefficient to eq.\,\ref{S21} gives the mode decay times $\bar{\tau}\simeq 65$\,ns of the hybridized system, compatible with the value of $\tau_2$ provided by the manufacturer and the measured $\tau_c$. 

\begin{figure}
\begin{center}
\includegraphics[width=3.3in]{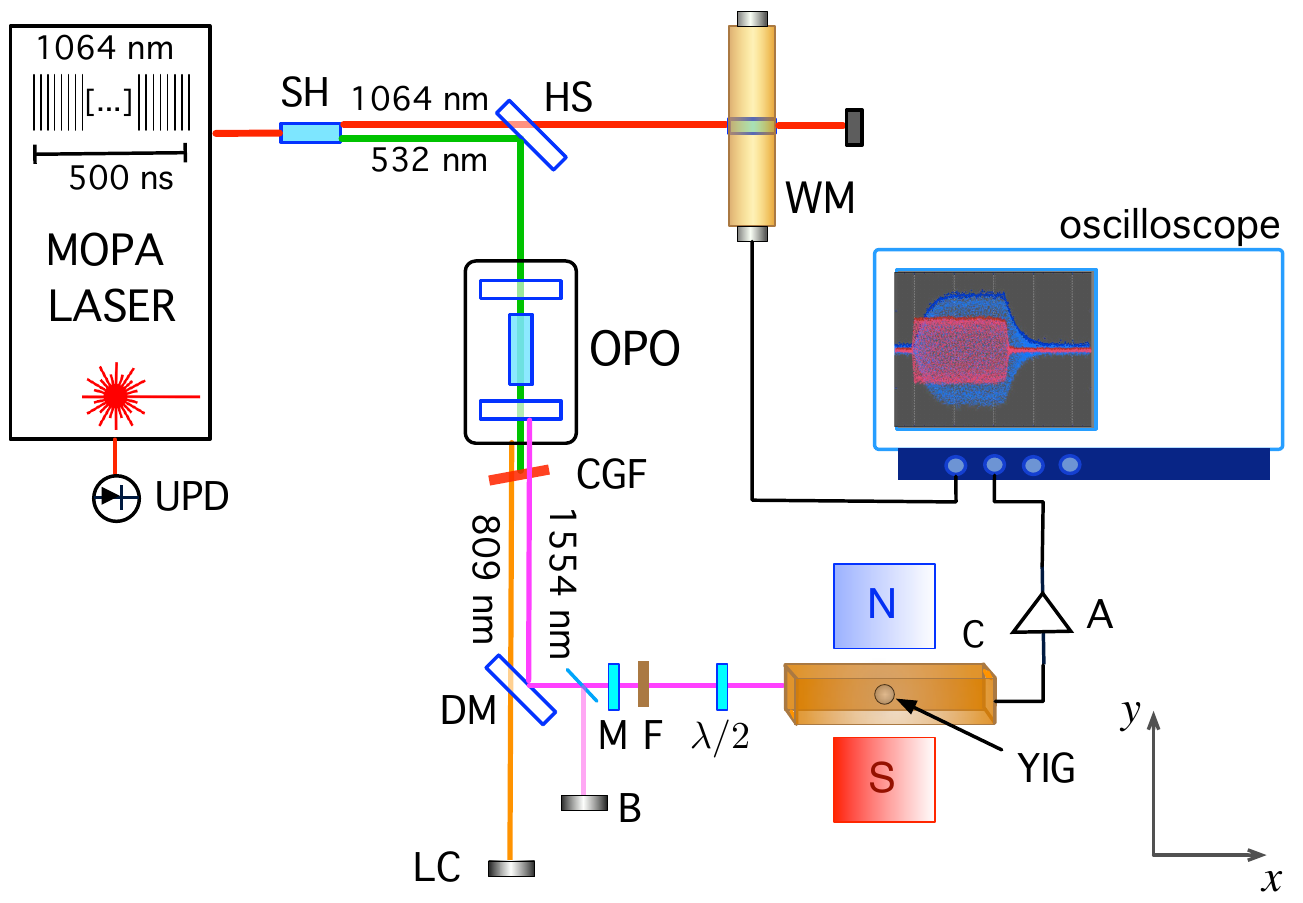}
\caption{Schematic representation of the experimental arrangement. The 1064\,nm--wavelength macro--pulse delivered by a MOPA laser is frequency--doubled (SH) to synchronously pump an optical parametric oscillator (OPO). The laser repetition rate, macro--pulse uniformity and  energy are monitored at an InGaAs ultrafast photodiode (UPD), a coaxial waveguide device WM \cite{braggio:2014} and bolometer B respectively. The 809\,nm OPO output beam intensity profile is adjusted at a digital laser camera LC. To make sure that only emission at 1550\,nm impinges on the YIG sphere, several optical filters are inserted in the beam path. CGF is a 610\,nm longpass colored glass filter, F transmits $\lambda>1500$\,nm and M is a 1064\,nm, high reflectivity dielectric mirror. HS (harmonic separator) is a  dielectric mirror that transmits 1064\,nm wavelength and has a high reflectivity for 532\,nm whereas DM is a 1000\,nm--cutoff wavelength dichroic mirror. The microwave field generated during the magnetization precession is detected by means of an antenna critically coupled to the TE$_{102}$ mode and connected through a short transmission line to a  39\,dB--gain amplification stage A. The amplified signal is finally registered at a 20\,GHz sampling oscilloscope.}
\label{sch}
\end{center}
\end{figure}

{\em Photoinduced magnetization.}---
Once the hybrid system has been characterized, the experimental apparatus illustrated in Fig.\,\ref{sch} is used to investigate the opto--magnetic phenomenon.
 The 7.2\,ps--duration, 1.55\,$\mu$m--wavelength laser pulses are obtained at the idler output of an optical parametric oscillator (OPO), synchronously pumped by the second harmonic of a MOPA (master oscillator power amplifier) laser system that has been described elsewhere \cite{Agnesi:08}. It is important to note that we are exploiting a non absorptive mechanism as the optical wavelength is within the YIG transparency window ($1.5-5\,\mu$m).   
The beam waist at the YIG position is $1.28$\,mm, and the average intensity of the incident pulses is $2.4$\,MW/cm$^2$, obtained within $<1\,\mu$s--duration macro--pulses.

\begin{figure}
\centering
\begin{minipage}{0.58\columnwidth}
  \centering
  \includegraphics[height=3.7 cm]{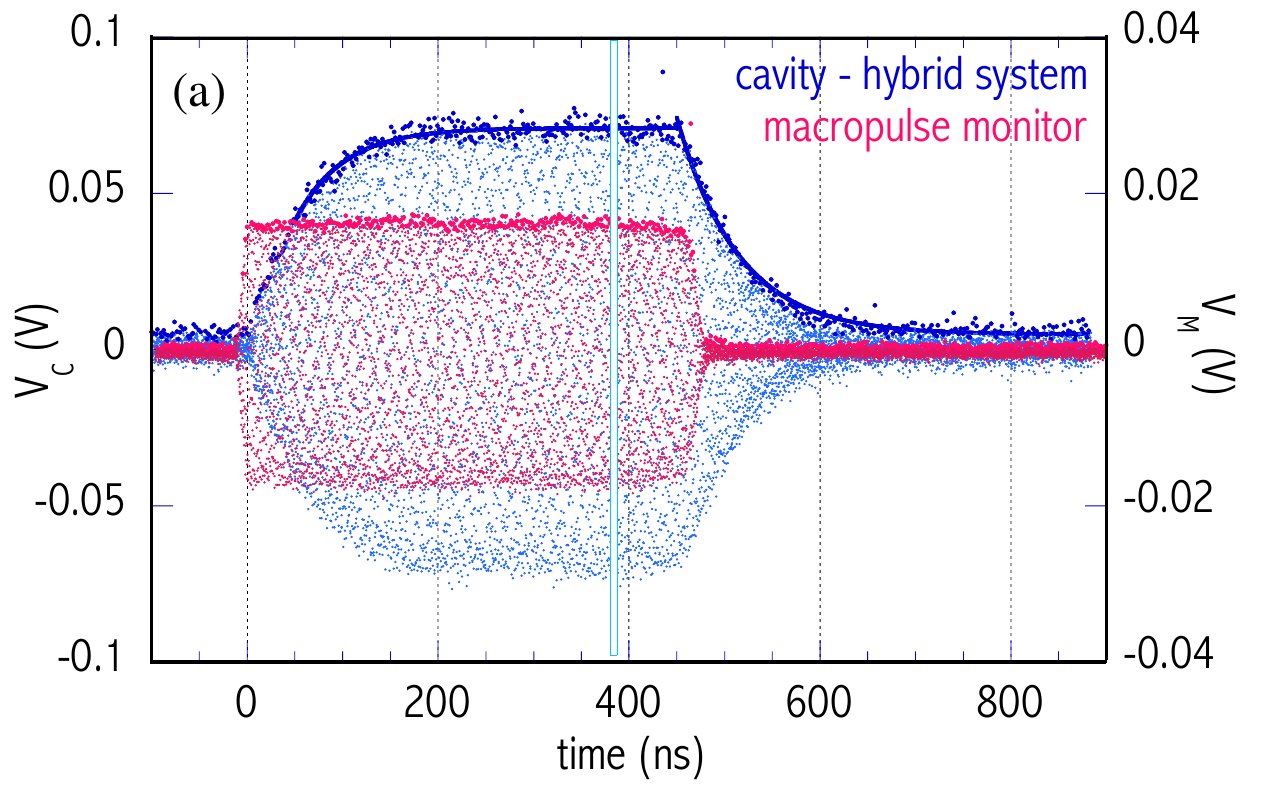}
\end{minipage}%
\begin{minipage}{0.55\columnwidth}
  \centering
 \includegraphics[width=2.8 cm ]{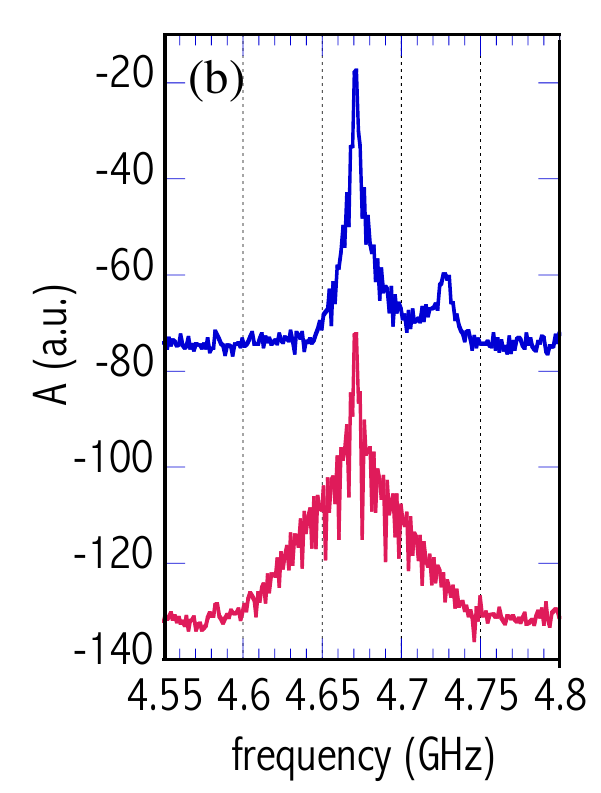}
  \label{fig:test2}
\end{minipage}
\begin{minipage}{0.5\columnwidth}
  \centering
  \includegraphics[height=3.6 cm]{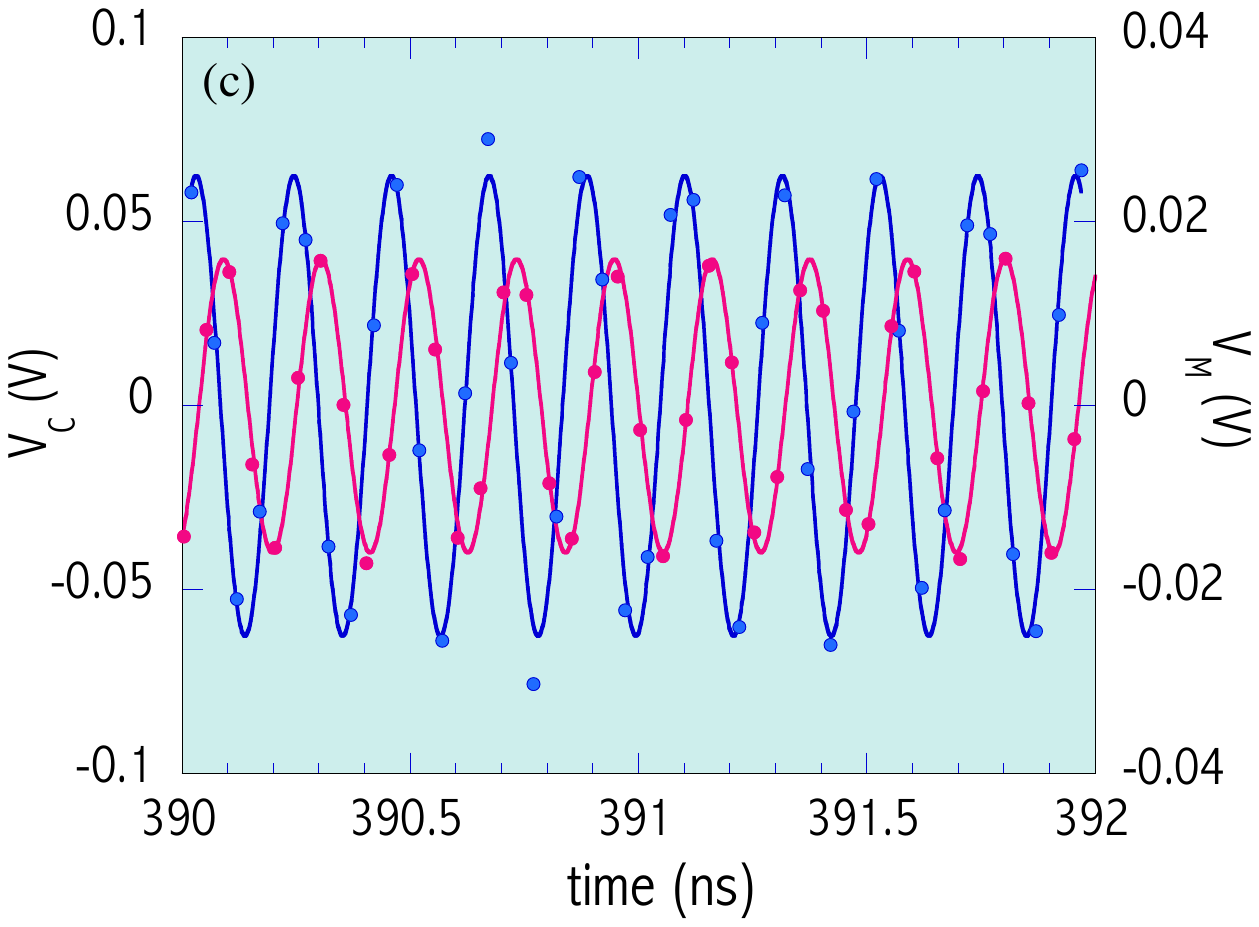}
\end{minipage}%
\begin{minipage}{0.58\columnwidth}
  \centering
 \includegraphics[width=3.6 cm ]{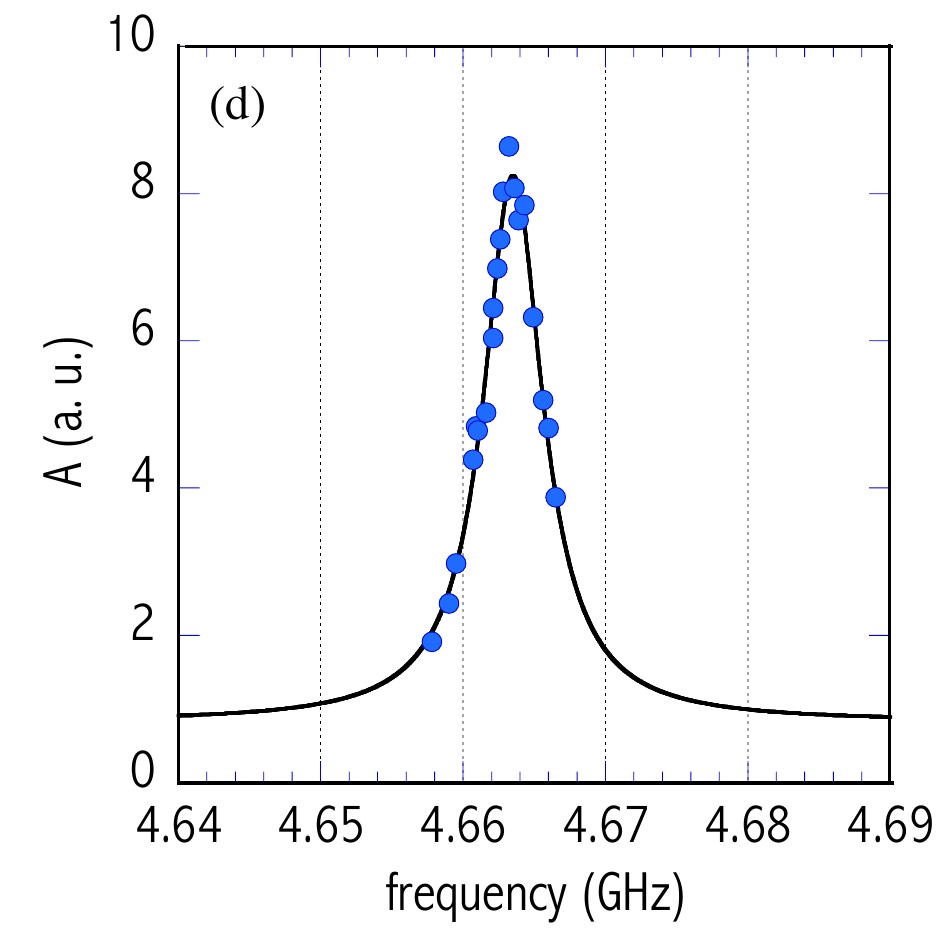}
\end{minipage}
  \caption{(Color online) Optically-driven spin precession in the time and frequency domain. (a)\,Oscilloscope traces displaying both the amplified signal V$_{\rm C}$ detected in the microwave cavity hosting the YIG sphere (blue) and the output V$_{\rm M}$ of the laser macro--pulse mo\-ni\-tor (red). (b)\,Fourier transform amplitude spectrum of the microwave signals displayed in (a). The logarithmic scale is used for the vertical axis.
  (c)\,  2\,ns--duration zoom out of (a) showing the magnetization precessing synchronously with the laser pulses.  
   (d)\, Tuning the laser repetition rate to the hybridized frequency $f_{-}$. \label{figu:3}}
   
\end{figure}

In Fig.\,\ref{figu:3} we demonstrate the all--optical coherent control of the magnon--photon mode at $f_{-}=4.67$\,GHz by employing a train of laser pulses with repetition rate $f_r$ tuned to $f_{-}$. 
The rise and decay time of the microwave pulse registered at the oscilloscope agrees with the mode decay time $\bar\tau=65\,$ns we get through the S$_{12}$ measurements within experimental errors. 
The duration of the optical excitation is set to a value of $t_e \simeq 0.5\,\mu{\rm s}  > \bar\tau$ allowing us to control the system in its steady state. This differs from previous studies in opto--magnetism  which were focused on the transient optical control of the magnetization via single femtosecond laser pulses \cite[see][and references therein]{Kirilyuk:2010}.
 Moreover, the YIG magnetization precesses in phase with the laser pulses, as demonstrated by juxtaposition in Fig.\,\ref{figu:3}\,(c) of the signal generated in the microwave cavity and the output of the laser macro--pulse monitor WM, i.e. a coaxial waveguide hosting a nonlinear crystal in which microwaves are generated through optical rectification \cite{braggio:2014}. 
 Another important signature of the coherent precession of the magnetization is also shown in Fig.\,\ref{figu:3}\,(d), where the amplitude of the Fourier transform of the microwave signal is plotted for different values of the laser repetition rate $f_{r}$. The data are fitted to a lorentzian curve that takes into account the convolution between the optical excitation and the profile of the hybridized mode at $4.6711$\,GHz.  As shown in Fig.\,\ref{figu:3}\,(b), the spectral component $f_{+}$ is also excited but with a much smaller strength. These results, combined with  the assessment of stationary precession of the macroscopic magnetization, unambiguously show that each macro--pulse acts as an effective microwave field on the ensemble of strongly correlated spins of the FMR mode.
 
\begin{figure}
\centering
\begin{minipage}{0.5\columnwidth}
  \centering
  \includegraphics[height=3.9  cm]{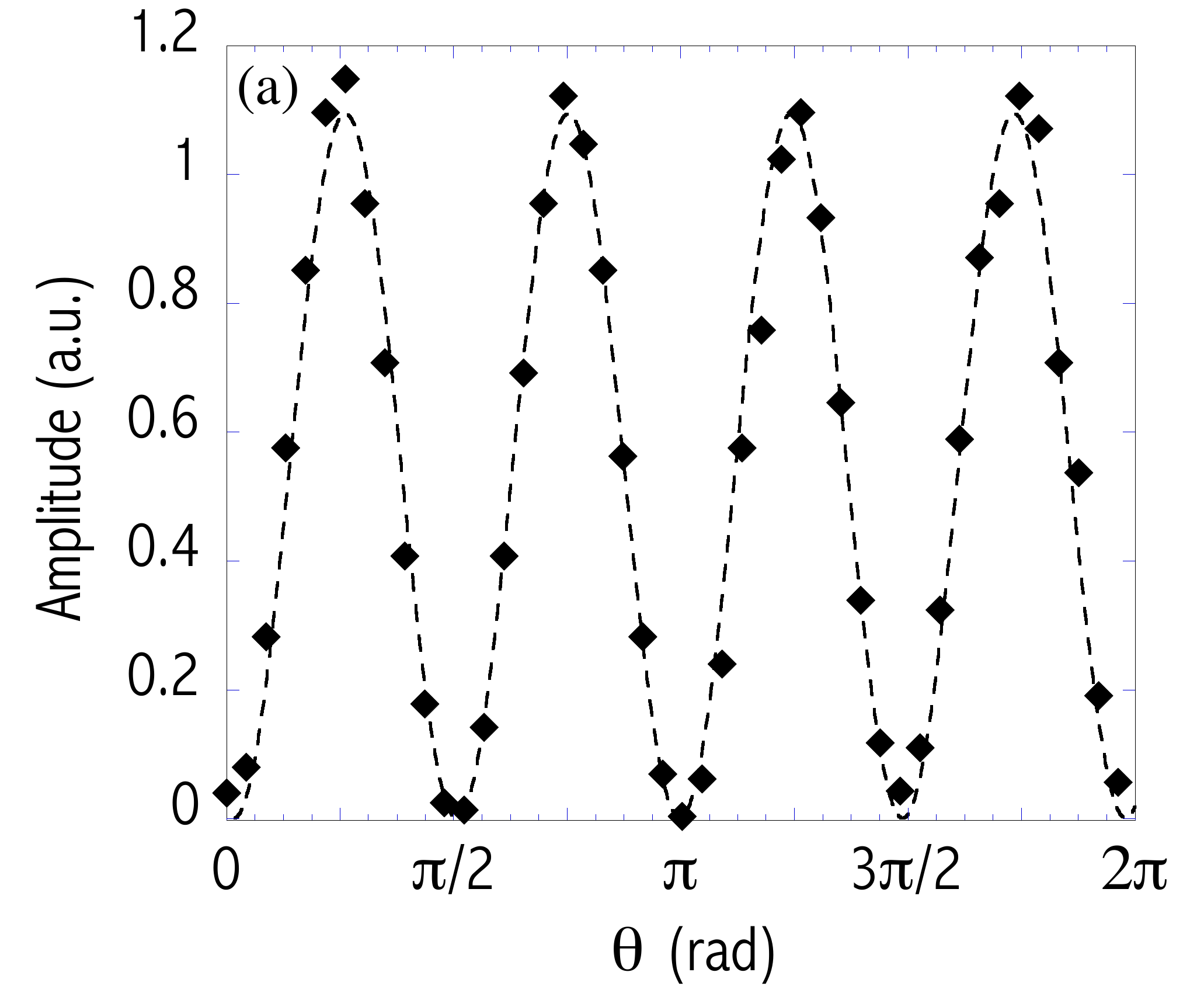}
\end{minipage}%
\begin{minipage}{0.5\columnwidth}
  \centering
  \includegraphics[height=3.85 cm]{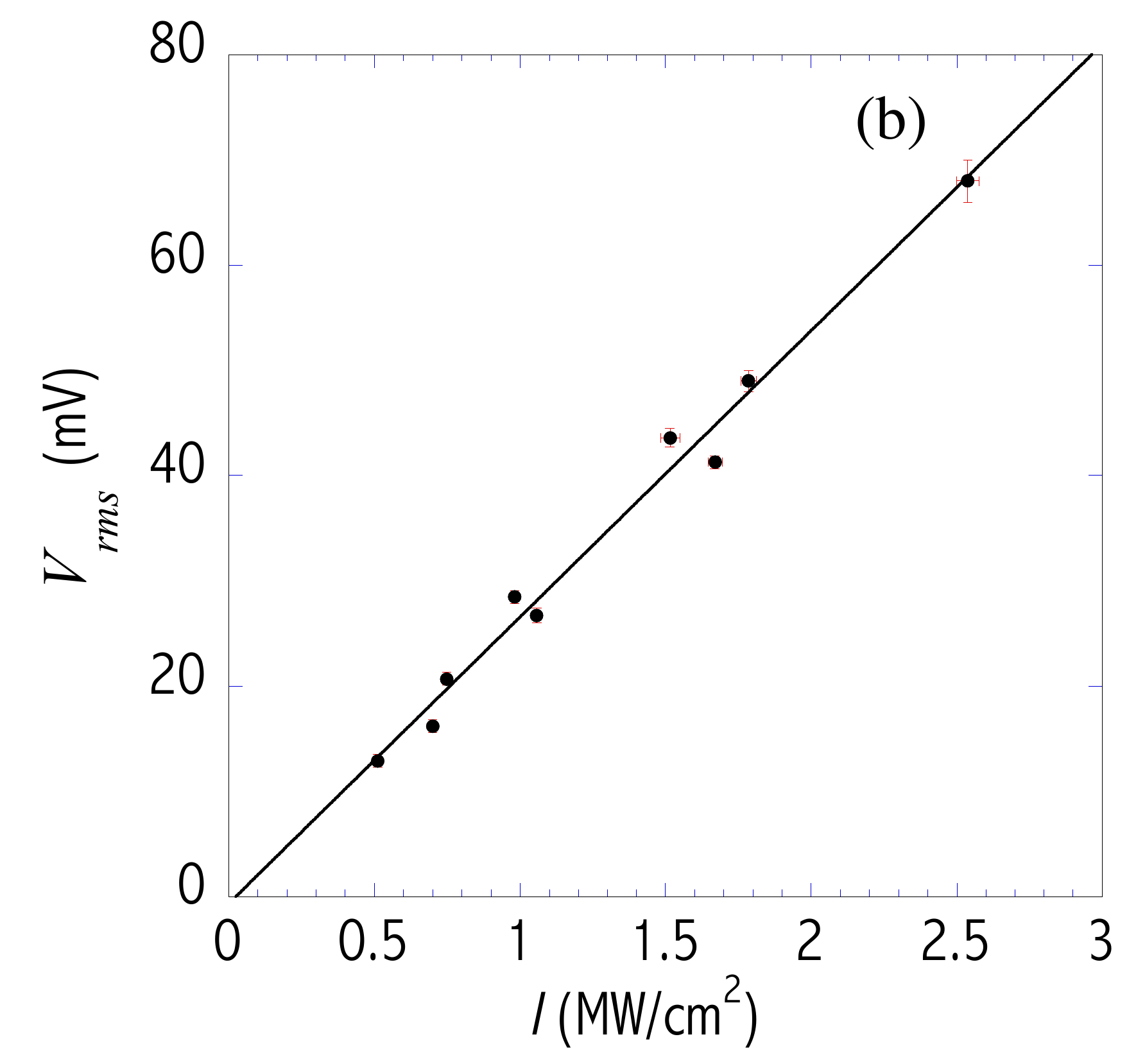}
\end{minipage}
  \caption{(a) Amplitude of the microwave signal in the cavity as a function of the laser polarization angle. (b) Magnetization dependence on the laser intensity. \label{fig:4}}
\end{figure}

{\em Discussion.}--- To confirm the nonthermal origin of  the laser--induced magnetization precession and definitely attribute the observed opto--magnetic phenomenon to the IF effect, we  investigated the dependence of the microwave signal amplitude on the laser polarization \cite{Kirilyuk:2010} and the results are reported in Fig.\,\ref{fig:4} (a). 
Owing to the strong anisotropy of  the magnetic susceptibility of YIG, the time--dependent magnetization vector $\mathbf{M}$ is not expected to be parallel to the wave vector $\mathbf{k}$ as is the case for IF in isotropic medium and circularly polarized light. 
Actually the observed magnetization precession is induced by a second--order process conveniently described by a third--rank, axial time--odd tensor $\chi^{(2)}_{ijk}$, provided that $\mathbf{k}$ is orthogonal to the $[1 \, 1\,  0]$ crystal direction 
$\mathbf{d}$ \cite{Kanda:2011}. 
As the reference system axes $x$ and $y$ coincide with  $\mathbf{k}$ and $\mathbf{d}$ directions, the photoinduced magnetization lies in the $yz$ plane and reads
\begin{equation}\label{mag}
M_i =\int d\omega \,  \chi^{(2)}_{ijk} E^\star_j(\omega)  E_k(\omega) \ , 
\end{equation}
where $i=2,3$, and $E_2(\omega)=E(\omega)\cos\varphi$ and $E_3(\omega)=E(\omega)\sin\varphi$; here $E(\omega)$ is the Fourier transform of the laser electric field and $ \varphi$ is the polarization angle of the incident light with respect to the $y$ axis. 
Due to well--known symmetries of second--order susceptibility \cite{Kirilyuk:2010}, the non--vanishing components of $\chi_{ijk}$ are $\chi_{233}=-\chi_{222}=\chi_{332}=\chi_{323} \equiv \Xi(\omega)$ and equation\,\ref{mag} gives the components
\begin{eqnarray}
\label{mag2}
M_{z} &=& \int d\omega \,  \Xi(\omega) \vert E(\omega) \vert ^2 \cos 2\varphi\\
M_{y}&=& \int d\omega \,  \Xi(\omega) \vert E(\omega) \vert ^2 \sin2\varphi .
\end{eqnarray}
Around the hybridized mode frequencies $\omega_{\pm}$, the real and imaginary part of the complex susceptibility $\Xi(\omega)$ can be approximated by absorption $\Xi(\omega)^{\prime \prime}$ and dispersion $\Xi(\omega)^{\prime}$ components of magnetization \cite{Fiorillo:2004}. In particular, at working frequency $\omega_{-}$ we have only absorption with no dispersion, hence the susceptivity $\Xi(\omega_-) = \Xi_0 \,\omega_{-} \bar\tau/2$  becomes real, and does not affect
 the magnetization direction. Thus the fulfillment of resonant condition also allow us to simplify the geometric description of the photoinduced magnetization vector.    
Indeed, to explain the 4--fold periodicity of the plot displayed in Fig.\,\ref{fig:4} (a) we only need to realize that the cavity selects  the $M_z\propto \cos 2\varphi $ component via its geometric projection on TE$_{102}$ mode (i.e. the z direction 
as shown in Fig. \ref{coup}), and that the critically coupled antenna cannot
 distinguish between parallel and antiparallel orientation of $M_{z}$. Therefore the 
detected magnetization signal must be proportional to $\vert \cos 2\varphi \vert$, as confirmed by the fit to the data in
Fig.\,\ref{fig:4} (a).  
Figure\,\ref{fig:4} (b) shows instead the linearity of the measured spin oscillations amplitude as a function of the pump laser intensity, in agreement with eq.\,\ref{mag}
 as well.

The strength of the effective microwave field $B_{\rm eff}$ that drives the $M_z$ precession can be estimated thanks to the peculiar dynamics of hybridization. In general, the  absorbed power in stationary conditions by a magnetized sample \cite{Fiorillo:2004} is given by 
\begin{equation} \nonumber
\label{pabs}
P_a  =  V_s \left \langle -\mathbf{B} \cdot \frac{d\mathbf{M}}{dt} \right \rangle \ , 
\end{equation}
where $\langle \cdot \rangle$ denotes the time average over one period. Moreover, at resonance and for a critically coupled inductive loop, the measured power in the microwave cavity is $ P_a/2$. In our experimental conditions, the absorbed power by the YIG crystal at the frequency $\omega_-$ 
\begin{equation}  
\label{pabsYIG}
P_a  =  V_s\, \Xi_0 \, \omega^2_- \, \overline{\tau} \, \frac{B_{\rm eff}^2}{\mu_0} \\ 
\end{equation}
 is written in terms of quantities that are measured or fitted to the data, so that the second--order susceptivity can be readily estimated through $ \Xi_0=P_a\mu_0/( V_s\, \omega_{-}^{2} \, \overline{\tau} B_{\rm eff}^2)$, where $B_{\rm eff}$ represents the laser induced effective magnetic field. Due to $1/f$ dependence of the power spectrum generated by downconversion of  the picosecond frequency comb, the infrared optical field average amplitude $B_l=\sqrt{\mu_0I/c}=10$\,mT, at $f_o\simeq 190$\,THz optical frequency, is suppressed to $B_{\rm eff}=  2.5 \times 10^{-5} B_I$ = 0.25\,$\mu$T at $f_-\sim 4.7$\,GHz.  With $P_a=3$\,nW  estimated 
  from the plots reported in Fig.\,\ref{figu:3}, we eventually get $\Xi_0 \sim 10^{-7}$ cm$^2$/MW. 

In summary, our experimental and theoretical approach provides a purely optical,  flexible technique to manipulate the magnetization vector in YIG via polarization rotation and intensity modulation of the incident laser beam. Remarkably, the maximum control speed of this process is only limited by the bandwidth of currently available electro-optic devices.  Unlike the ingenious optical method described in reference \cite{Kanda:2011}, here the mode-locked pulses impinging on the magnetized material allow for operation of the system in the steady state, opening a path on the ultrafast laser control of hybridized magnon-photon systems. It is worth mentioning that commercially available compact ultrafast oscillators with 200\,pJ-energy output pulses \cite{krainer:2002} may foster applications of the present method in the opto--magnetism field.

The authors thank D. Budker and V. S. Zapasskii for carefully reading the manuscript and for useful discussions. C.\,B. and M.\,G. acknowledge partial financial support of the University of Padova under Progetto di Ateneo (reference grant number CPDA135499/13).  
Technical support by E. Berto is gratefully acknowledged.

\end{document}